\documentstyle[twoside,fleqn,espcrc2,graphicx,rotate,epsfig]{article}
\newcommand{\be}{\begin{equation}}
\newcommand{\ee}{\end{equation}}

\title{ 
\vspace{-2cm}
\hfill \rm \null \hfill
 \hbox{\normalsize ADP-01-54/T486} \\
\vspace{-2mm}
 \hfill \hbox{\normalsize JLAB-THY-01-41} \\
\vspace{-1mm}
Novel fat-link fermion actions 
}

\author{
J.~M.~Zanotti\address[CSSM]{Special Research Center for the
	Subatomic Structure of Matter, and		\\
	Department of Physics and Mathematical Physics,
	Adelaide University, 5005, Australia},
S.~Bilson-Thompson\addressmark[CSSM],
F.~D.~R.~Bonnet\addressmark[CSSM],
P.~D.~Coddington\addressmark[CSSM],
D.~B.~Leinweber\addressmark[CSSM],
A.~G.~Williams\addressmark[CSSM],
J.~B.~Zhang\addressmark[CSSM],
W.~Melnitchouk\addressmark[CSSM]$^,$\address[JLab]{Jefferson Lab, 12000
	Jefferson Avenue, Newport News, VA 23606, U.S.A.}and
F.~X.~Lee\addressmark[JLab]$^,$\address{Center for Nuclear Studies,
	Department of Physics, \\
	The George Washington University, Washington, D.C. 20052, U.S.A.}}

\begin{document}

\thispagestyle{empty}

\begin{abstract}
The hadron mass spectrum is calculated in lattice QCD using a novel
fat-link clover fermion action in which only the irrelevant operators
of the fermion action are constructed using smeared links.  The
simulations are performed on a $16^3 \times 32$ lattice with a lattice
spacing of $a=0.125$ fm.  We compare actions with $n=4$ and 12
smearing sweeps with a smearing fraction of 0.7. The $n=4$ Fat Link
Irrelevant Clover (FLIC) action provides scaling which is superior to
mean-field improvement, and offers advantages over nonperturbative
${\cal O}(a)$ improvement.
\end{abstract}

\maketitle

%\vspace{3mm}
%PACS number(s): 11.15.Ha, 12.38.Gc, 12.38.Aw

%\newpage

%%%%%%%%%%%%%%%%%%%%%%%%%%%%%%%%%%%%%%%%%%%%%%%%%%%%%%%%%%%%%%%%%%%%%%%%%%%
\section{INTRODUCTION}

Understanding the generation of hadron mass from first principles has proved
to be challenging.
The only method for deriving hadron masses directly
from QCD is a numerical calculation on the lattice.
The high computational cost required to perform accurate lattice
calculations at small lattice spacings, however, has led to an
increased interest in quark action improvement.
To avoid the famous doubling problem, Wilson fermions \cite{Wilson}
introduce additional terms which explicitly break chiral symmetry at
${\cal O}(a)$.  To extrapolate reliably to the continuum, simulations
must be performed on fine lattices, which are very computationally
expensive.  The scaling properties of the Wilson action at finite $a$
can be improved by introducing any number of irrelevant operators of
increasing dimension which vanish in the continuum limit.

The Sheikholeslami-Wohlert (clover) action \cite{Sheikholeslami:1985ij}
introduces an additional irrelevant dimension-five operator to the
standard Wilson \cite{Wilson} quark action, 
\be
S_{\rm SW} = S_{\rm W} - \frac{i a C_{\rm SW} r}{4}\
	 \bar{\psi}(x)\sigma_{\mu\nu}F_{\mu\nu}\psi(x)\ ,
\label{clover}
\ee
where $S_{\rm W}$ is the standard Wilson action
and $C_{\scriptstyle {\rm SW}}$ is the clover coefficient which can be tuned to
remove ${\cal O}(a)$ artifacts.
Nonperturbative (NP) ${\cal O}(a)$ improvement \cite{Luscher:1996sc}
tunes $C_{\scriptstyle {\rm SW}}$ to all powers in $g^2$ and displays
excellent scaling, as shown by Edwards {\it et
al}. \cite{Edwards:1998nh}, who studied
the scaling properties of the nucleon and vector meson masses for various
lattice spacings (see also Section~\ref{discussion} below).
In particular, the linear behavior of the NP-improved clover actions,
when plotted against $a^2$, demonstrates that ${\cal O}(a)$ errors are
removed.  It was also found in Ref.~\cite{Edwards:1998nh} that a
linear extrapolation of the mean-field improved data fails, indicating
that ${\cal O}(a)$ errors are still present.

A drawback to the clover action, however, is the associated problem of
exceptional configurations, where the quark propagator encounters
singular behavior as the quark mass becomes small.  In practice, this
prevents the use of coarse lattices ($\beta < 5.7 \sim a > 0.18$~fm)
\cite{Bardeen:1998gv,DeGrand:1998jq}.
Furthermore, the plaquette version of $F_{\mu\nu}$, which is commonly used in
Eq.~(\ref{clover}), has large ${\cal O}(a^2)$ errors, which can lead
to errors of the order of $10 \%$ in the topological charge even on
very smooth configurations \cite{Bonnet:2000dc}.

The idea of using fat links in fermion actions was first explored by
the MIT group \cite{Chu:1994vi} and more recently has been studied by
DeGrand {\it et al}.
\cite{DeGrand:1998jq,DeGrand:1999gp}, who showed that the exceptional
configuration problem can be overcome by using a fat-link (FL) clover
action.  Moreover, the renormalization of the coefficients of action
improvement terms are small.
A drawback to conventional fat-link techniques, however, is that in
smearing the links 
gluon interactions are removed at the scale of the cutoff.  While this
has some tremendous benefits, the short-distance quark interactions
are lost. As a result decay constants, which are sensitive to the wave
function at the origin, are suppressed.

A solution to these problems is to work with two sets of links in the
fermion action.  In the relevant dimension-four operators, one works
with the untouched links generated via Monte Carlo methods, while the
smeared fat links are introduced only in the higher dimension
irrelevant operators.  In this way the continuum limit of the theory
is perfectly well defined.

% In this talk I will present the first results of simulations of the
In this paper we present the first results of simulations of the
spectrum of light mesons and baryons using this variation on the
clover action.
In particular, we will start with the standard clover action and
replace the links in the irrelevant operators with APE smeared
\cite{ape}, or fat links.
We shall refer to this action as the Fat-Link Irrelevant Clover (FLIC)
action.  Although the idea of using fat links only in the irrelevant
operators of the fermion action was developed here independently,
suggestions have appeared previously \cite{neub}. To the best of our
knowledge, this is the first report of lattice QCD calculations using
this novel fermion action.

\vspace*{-0.1cm}
%%%%%%%%%%%%%%%%%%%%%%%%%%%%%%%%%%%%%%%%%%%%%%%%%%%%%%%%%%%%%%%%%%%%%%%%%%%
\section{GAUGE ACTION}
\label{simulations}

The simulations are performed using a tree-level
${\cal O}(a^2)$--Symanzik-improved 
\cite{Symanzik:1983dc} gauge action on a $16^3 \times 32$ lattice at
$\beta = 4.60$, providing a lattice spacing $a=0.125(2)$~fm
determined from the string tension with $\sqrt\sigma=440$~MeV.
A total of 50 configurations are used in this analysis, and the error
analysis is performed by a third-order, single-elimination jackknife,
with the $\chi^2$ per degree of freedom ($N_{\rm DF}$) obtained via
covariance matrix fits.

%%%%%%%%%%%%%%%%%%%%%%%%%%%%%%%%%%%%%%%%%%%%%%%%%%%%%%%%%%%%%%%%%%%%%%%%%%%
\section{FAT-LINK IRRELEVANT FERMION ACTION}
\label{FLinks}

Fat links \cite{DeGrand:1998jq,DeGrand:1999gp} are created by averaging
or smearing links on the lattice with their nearest neighbors in a
gauge covariant manner (APE smearing).
The smearing procedure \cite{ape} replaces a link, $U_{\mu}(x)$, with a
sum of the link and $\alpha$ times its staples
\begin{eqnarray}
U_{\mu}(x)\ &\rightarrow&\ U_\mu^{\rm FL}(x)\ =\
(1-\alpha) U_{\mu}(x) \\
&&\hspace*{-12mm} + \frac{\alpha}{6}\sum_{\nu=1 \atop \nu\neq\mu}^{4}
  \Big[	U_{\nu}(x)
	U_{\mu}(x+\nu a)
	U_{\nu}^{\dag}(x+\mu a)				\nonumber \\
\mbox{} 
&&\hspace*{-12mm} + U_{\nu}^{\dag}(x-\nu a)
	U_{\mu}(x-\nu a)
	U_{\nu}(x-\nu a +\mu a)
  \Big] \, , \nonumber
\end{eqnarray} 
followed by projection back to SU(3). We select the unitary matrix
$U_{\mu}^{\rm FL}$ which maximizes
$$
{\cal R}e \, {\rm{tr}}(U_{\mu}^{\rm FL}\, U_{\mu}'^{\dagger})\,
$$
by iterating over the three diagonal SU(2) subgroups of SU(3).
We repeat this procedure of smearing followed immediately by projection $n$ times.
We create our fat links by setting $\alpha = 0.7$ and comparing $n=4$ and 12 smearing
sweeps.
The mean-field improved FLIC action now becomes
\be
S_{\rm SW}^{\rm FL}
= S_{\rm W}^{\rm FL} - \frac{iC_{\rm SW} \kappa r}{2(u_{0}^{\rm FL})^4}\
	     \bar{\psi}(x)\sigma_{\mu\nu}F_{\mu\nu}\psi(x)\ ,
\ee
where $F_{\mu\nu}$ is constructed using fat links, and where the
mean-field improved Fat-Link Irrelevant Wilson action is
\begin{eqnarray}
S_{\rm W}^{\rm FL}
 &=&  \sum_x \bar{\psi}(x)\psi(x) \\
&+& \kappa \sum_{x,\mu} \bar{\psi}(x)
    \bigg[ \gamma_{\mu}
      \bigg( \frac{U_{\mu}(x)}{u_0} \psi(x+\hat{\mu}) \nonumber\\
& & \qquad - \frac{U^{\dagger}_{\mu}(x-\hat{\mu})}{u_0} \psi(x-\hat{\mu})
      \bigg)						\nonumber\\
&-& r \bigg(
	  \frac{U_{\mu}^{\rm FL}(x)}{u_0^{\rm  FL}} \psi(x+\hat{\mu}) \nonumber\\
& & \qquad + \frac{U^{{\rm FL}\dagger}_{\mu}(x-\hat{\mu})}{u_0^{\rm FL}}
	  \psi(x-\hat{\mu})
      \bigg)
    \bigg]\ , \nonumber
\end{eqnarray}
with $\kappa = 1/(2m+8r)$. We take the standard value $r=1$.
The $\gamma$-matrices are hermitian and
$\sigma_{\mu\nu} = [\gamma_{\mu},\ \gamma_{\nu}]/(2i)$.

As reported in Table~\ref{meanlink}, the mean-field improvement
parameter for the fat links is very close to 1.  Hence, the mean-field
improved coefficient for $C_{\rm SW}$ is expected to be
adequate{\footnote{Our experience with topological charge operators
suggests that it is advantageous to include $u_0$ factors, even as
they approach 1.}}.  In addition, actions with many irrelevant
operators ({\it e.g.} the D$_{234}$ action) can now be handled with
confidence as tree-level knowledge of the improvement coefficients
should be sufficient.  Another advantage is that one can now use
highly improved definitions of $F_{\mu\nu}$ (involving terms up to
$u_0^{12})$, which give impressive near-integer results for the
topological charge \cite{sbilson}.

In particular, we employ an ${\cal O}(a^4)$ improved definition of
$F_{\mu\nu}$ \cite{sbilson} in which the standard clover-sum of four
$1 \times 1$ Wilson loops lying in the $\mu ,\nu$ plane is combined
with $2 \times 2$ and $3 \times 3$ Wilson loop clovers.

\begin{table}
\begin{center}
\caption{The value of the mean link for different numbers of smearing
	sweeps, $n$.\label{meanlink}}
\vspace*{0.5cm}
\begin{tabular}{ccc}
\hline \vspace*{0.9mm}
$n$ & $u^{\rm FL}_0$ & $(u^{\rm FL}_0)^4$ \\ \hline
    0  &  0.88894473 & 0.62445197 \\
    4  &  0.99658530 & 0.98641100 \\
    12 &  0.99927343 & 0.99709689 \\
\hline
\end{tabular}
%\vspace*{0.5cm}
\end{center}
\end{table}

% \newpage

Work by DeForcrand {\it et al}. \cite{deForcrand:1997sq} suggests that
7 cooling sweeps are required to approach topological charge within
1$\%$ of integer value.  This is approximately 16 APE smearing sweeps
at $\alpha = 0.7$ \cite{Bonnet:2001rc}.  However, achieving integer
topological charge is not necessary for the purposes of studying
hadron masses, as has been well established.  To reach integer
topological charge, even with improved definitions of the topological
charge operator, requires significant smoothing and associated loss of
short-distance information.  Instead, we regard this as an upper limit
on the number of smearing sweeps.

Using unimproved gauge fields and an unimproved topological charge
operator, Bonnet {\it et al}. \cite{Bonnet:2000dc} found that the
topological charge settles down after about 10 sweeps of APE smearing
at $\alpha=0.7$.  Consequently, we create fat links with APE smearing
parameters $n=12$ and $\alpha=0.7$.  This corresponds to $\sim 2.5$
times the smearing used in Refs.~\cite{DeGrand:1998jq,DeGrand:1999gp}.
Further investigation reveals that improved gauge fields with a small
lattice spacing ($a=0.125$~fm) are smooth after only 4 sweeps. Hence,
we perform calculations with 4 sweeps of smearing at $\alpha=0.7$ and
consider $n=12$ as a second reference.
Table~\ref{meanlink} lists the values of $u_0^{\rm FL}$
for $n=0$, 4 and 12 smearing sweeps.

A fixed boundary condition is used for the fermions by setting
\be
U_t (\vec{x},nt) = 0 \,\,\, {\rm and}\,\,\, U_t^{\rm FL} (\vec{x},nt) = 0 
\qquad \forall\ \vec{x}\ 
\ee
in the hopping terms of the fermion action.  The fermion source is
centered at the space-time location {$(x,y,z,t) = (1,1,1,3)$}, which
allows for two steps backward in time without loss of signal.
Gauge-invariant gaussian smearing \cite{Gusken:qx} in the spatial
dimensions is applied at the source to increase the overlap of the
interpolating operators with the ground states.

\section{RESULTS}
\label{discussion}

Hadron masses are extracted from the Euclidean time dependence of the
calculated two-point correlation functions. 
The effective masses are given by
\be
M(t+1/2) = \log [G(t)] - \log[G(t+1)]\ .
\label{effmass}
\ee
The critical value of $\kappa$, $\kappa_{c}$, is determined by linearly
extrapolating $m_{\pi}^2$ as a function of $m_q$ to zero. 
We used five values of quark mass and the strange quark mass was taken
to be the second heaviest quark mass.

Effective masses~(\ref{effmass}) are calculated as a function of time
and various time-fitting intervals are tested with a covariance matrix
to obtain $\chi^2 / N_{\rm DF}$.
Good values of $\chi^2 / N_{\rm DF}$ are obtained for many different
time-fitting intervals as long as one fits after time slice 8.  All
fits for this action (``FLIC4'') are therefore performed on time slices
9 through 14.  For the Wilson action and the FLIC action with $n=12$
(``FLIC12'') the fitting regimes used are 9-13 and 9-14, respectively.

\begin{figure}[t]
\begin{center}
\epsfysize=7.5truecm
\leavevmode
\rotate[l]{\epsfbox{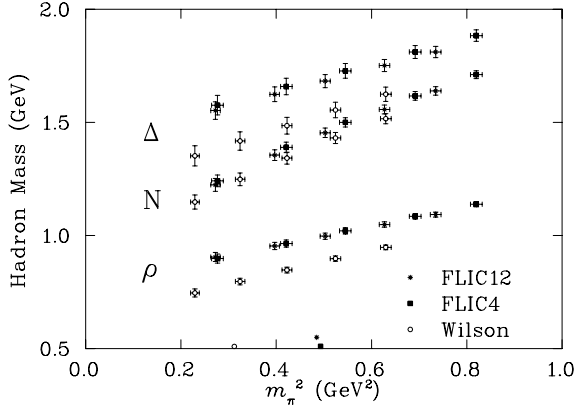} }
\vspace*{-1.0cm}
\caption{Masses of the nucleon, $\Delta$ and $\rho$ meson versus
	$m_{\pi}^2$ for the FLIC4, FLIC12 and Wilson actions.}
\label{Mvsmpi2}
\end{center}
\end{figure}

The behavior of the $\rho$, nucleon and $\Delta$ masses as a function
of squared pion mass is shown in Fig.~\ref{Mvsmpi2} for the various
actions.  The first feature to note is the excellent agreement between
the FLIC4 and FLIC12 actions.  On the other hand, the Wilson action
appears to lie somewhat low in comparison.  It is also reassuring that
all actions give the correct mass ordering in the spectrum.  The value
of the squared pion mass at $m_{\pi}/m_{\rho} = 0.7$ is plotted on the
abscissa for the three actions as a reference point.  This point is
chosen in order to allow comparison of different results by
{interpolating them to a common value of $m_{\pi}/m_{\rho} = 0.7$,
rather than extrapolating them to smaller quark masses, which is
subject to larger systematic and statistical uncertainties.

The scaling behavior of the different actions is illustrated in
Fig.~\ref{scaling1}.  The present results for the Wilson action agree
with those of Ref.~\cite{Edwards:1998nh}.
The first feature to observe in Fig.~\ref{scaling1} is that actions
with fat-link irrelevant operators perform extremely well.  For both
the vector meson and the nucleon, the FLIC4 action performs
systematically better than the FLIC12.  This suggests that 12 smearing
sweeps removes too much short-distance information from the
gauge-field configurations.  On the other hand, 4 sweeps of smearing
combined with our ${\cal O} (a^4)$ improved $F_{\mu\nu}$ provides
excellent results, without the fine tuning of $C_{\rm SW}$ in the NP
improvement program.  Notice that for the $\rho$ meson, a linear
extrapolation of the mean-field improved clover points in
Fig.~\ref{scaling1} indicates that there is better improvement when
using fat links in the irrelevant operators.
While there are no NP-improved clover plus improved glue simulation
results at $a^2 \sigma \sim 0.08$,
the simulation results that are available indicate that the fat-link
results also compete well with those obtained with a NP-improved
clover fermion action.

\begin{figure}[t]
\begin{center}
\epsfysize=7.5truecm
\leavevmode
\rotate[l]{\epsfbox{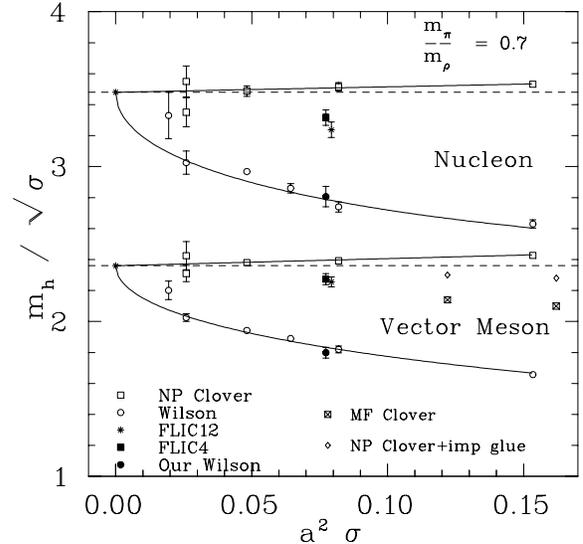} }
\vspace*{-1.0cm}
\caption{Nucleon and vector meson masses for the Wilson, NP-improved
	and FLIC actions obtained by interpolating our results
        of Fig.~{\protect \ref{Mvsmpi2}} to $m_\pi / m_\rho = 0.7$.
	Results from the present simulations are indicated by the
	solid points.
	The fat links are constructed with $n=4$ (solid squares) and
	$n=12$ (stars) smearing sweeps at $\alpha = 0.7$.}
\label{scaling1}
\end{center}
\end{figure}

Finally, we compare the convergence rates of the different actions
by comparing the number of Stabilised Biconjugate Gradient (BiCGStab) 
\cite{BiCG} iterations required to invert the fermion matrix.
The FLIC12 action converges at a slightly faster rate to the Wilson
fermion action and the FLIC4 action is the clear winner due to the
small number of BiCGStab iterations required to invert the fermion
matrix.  This provides great promise for performing simulations at
quark masses closer to the physical values.

%%%%%%%%%%%%%%%%%%%%%%%%%%%%%%%%%%%%%%%%%%%%%%%%%%%%%%%%%%%%%%%%%%%%%%%%%%%
% \section{Conclusions and Future Work}
\section{CONCLUSIONS}
\label{conclusion}

We have examined the hadron mass spectrum using a novel Fat Link
Irrelevant Clover (FLIC) fermion action, in which only the irrelevant,
higher-dimension operators involve smeared links.  One of the main
conclusions of this work is that the use of fat links in the
irrelevant operators provides excellent results. Fat links promise
improved scaling behavior over mean-field improvement. This technique
also solves a significant problem with ${\cal O}(a)$ nonperturbative
improvement on mean field-improved gluon configurations. Simulations
are possible and the results are competitive with
nonperturbative-improved clover results on plaquette-action gluon
configurations.
We have found that minimal smearing holds the promise of better
scaling behavior.  Our results suggest that too much smearing removes
relevant information from the gauge fields, leading to poorer
performance.  Fermion matrix inversion for FLIC4 is more efficient and
results show no sign of exceptional configuration problems.

This work paves the way for promising future studies.  It will be of
great interest to consider different lattice spacings to further test
the scaling of the fat-link actions.  Furthermore, the exceptional
configuration issue can be explored by pushing the quark mass down to
lower values.  The $n=4$ FLIC action holds great promise for
circumventing this issue as evidenced by the relative ease with which
one can invert the fermion matrix.  A study of the spectrum of excited
hadrons using the fat-link clover actions is currently in progress
\cite{Nstar}.

%%%%%%%%%%%%%%%%%%%%%%%%%%%%%%%%%%%%%%%%%%%%%%%%%%%%%%%%%%%%%%%%%%%%%%%%%%%
%\acknowledgements
\vspace*{0.3cm}
This work was supported by the Australian Research Council.  We would
also like to thank the National Computing Facility for Lattice Gauge
Theories for the use of the Orion Supercomputer.  W.M. and F.X.L. were
partially supported by the U.S. Department of Energy contract
\mbox{DE-AC05-84ER40150}, under which the Southeastern Universities
Research Association (SURA) operates the Thomas Jefferson National
Accelerator Facility (Jefferson Lab).

%%%%%%%%%%%%%%%%%%%%%%%%%%%%%%%%%%%%%%%%%%%%%%%%%%%%%%%%%%%%%%%%%%%%%%%%%%%

\end{document}